\documentclass[11pt]{article}
\usepackage{mathtools}
\usepackage{amssymb}
\usepackage{bm}
\usepackage[all, cmtip]{xy}
\begin{document}
	\title{Covariant phase space quantization of cosmological models}
	\author{Weixuan Hu}
	\date{}
	\maketitle	
\begin{abstract}

Covariant phase space quantization attempts to quantize the full space of classical solutions, leading to a quantum theory in which the usual time coordinate is missing.  In this paper we explore how the time evolution of the quantum states of this system is restored. We compare covariant phase space methods for
a nonrelativistic system (the harmonic oscillator) and three
relativistic quantum cosmologies. We find that for the nonrelativistic case
the preferred time coordinate emerges in a phase factor
when one transforms among certain
representations, giving exactly the phase that appears when the time evolution operator acts on the stationary states in the usual quantization method.  For the relativistic cases, on the other hand, this phase factor is
no longer present, and covariance appears to be restored. These results are consistent with other quantization methods for these models.

\end{abstract}

\section{Introduction}
Covariant phase space\footnote{A rich list of references about the covariant phase space can be found in \cite{ashtekar1991covariant}. } is the space of solutions of the equations of motions equipped with the covariant symplectic 2-form, the Lagrange form \cite{souriau1997structure}. When the space of solutions of a dynamical system\footnote{Such dynamical systems exists e.g. those with a regular Lagrangian (non-singular0 Lagrangian and complete Hamiltonian vector fields \cite{woodhouse1997geometric}.} is uniquely determined by the values of solutions at some specific moment, these values can be treated as a set of coordinates for the manifold of the space of solutions, and the classical time evolution of the phase space, in the language of covariant phase space, is described as the transformation of these coordinates   \cite{woodhouse1997geometric}. In this article, we explore this feature of the covariant phase space in the quantum aspect, especially the role of the time parameter $t$ during the quantization process. 

In a non-relativistic dynamical system in quantum mechanics  \cite{shankar1994principles}, the time parameter $t$ is an external variable of the system, and the time evolution of the state is described by an overall phase $e^{-iEt}$ that evolves as time progresses, where $E$ is an eigenvalue of Hamiltonian operator $\hat{\mathcal{H}}$, the generator of time translations.  When the covariant phase space of this dynamical system is quantized, the time evolution of the quantum state is not so obvious, as the covariant phase space is time independent: each point represents the entire history of a dynamical trajectory of the system, so there is no time parameter $t$ involved from the very beginning. 

In this article we show that the time evolution of quantum states of the covariant phase space may be restored by considering the change of a state among different representations. Here by a representation we mean a basis of the Hilbert space formed by a set of eigenstates of the ``coordinate operators.''  As noted above, the manifold of the space of solutions can be  uniquely specified by the values of the solutions at a particular moment.  Thus these values serve as the coordinates of this manifold. The ``coordinate operators'' are these coordinates considered as operators. To obtain a specific representation, we first pick  a set of coordinates for the space of solutions, and then treat these coordinates as operators, using their eigenstates to construct a basis of the Hilbert space, which is the representation we want. The quantum stats can be transformed to a different representation by choosing another set of ``coordinate operators'' to obtain another basis for the Hilbert space.

In Section $4$ we will see that when the quantum states are transformed among different representations, an extra phase that includes the time parameter $t$ will appear in  the new representation.  This time dependent phase is exactly the $e^{-iEt}$  phase that describes the time evolution in other approaches to quantum mechanics \cite{shankar1994principles}. This process of restoring time evolution by changing representations of the Hilbert space seems to be a quantum analogy of the classical dynamics in covariant phase space ,where the classical time evolution of the system is realized by the transformation of coordinates of the space of solutions.

In relativistic models, the time parameter $t$ is an intrinsic variable and Hamiltonian is a constraint. It is interesting to see how $t$ is involved in the quantization of their covariant phase space. In Section 4 we quantize the covariant phase space of three cosmological models, an FRW Universe coupled with dilaton field \cite{PhysRevD.77.044023}, an FRW Universe with cosmological constant \cite{Bina:2007wj}, and a Bianchi Type I Universe \cite{2007}. To explore the time dependence of the quantum states of these models, we switch from one Hilbert space representation to another as in the non-relativistic model. Our results show that in these relativistic models no time dependent phase emerges. This is consistent with other quantization method applied to these models.  A time dependent phase indicates the time evolution of the quantum states, while in relativistic models the time parameter $t$ is artificial, and can be chosen rather arbitrarily, so the quantum states of these models should not prefer a particular time in their evolution.

This article is organized as follows: in Section $2$ a brief review of the covariant phase space, including the construction of the covariant symplectic form, is given. In Section $3$ a detailed process for quantizing the covariant phase space is exhibited. In Section $4$ this quantization method is applied to a non-relativistic model (the harmonic oscillator) and three relativistic cosmological models. In Section $5$ we give a summary of the major idea of this article and some discussion of the conclusions.  As the quantization method of the covariant phase space proposed here is rather tentative and has only been tested on some simple examples, the conclusions may only apply to simple models with restricted conditions. A more formal canonical formalism developed in the covariant phase space of field theories can be found in \cite{ashtekar1991covariant, 1987thyg.book..676C, doi:10.1063/1.528801}, quantization of other 2D gravity model  in \cite{NAVARROSALAS1993291}, the covariant phase space formalism for cosmological models with infinite number of degrees of freedom in \cite{Torre:2002xt}, a more recent comprehensive discussion about the covariant phase space with boundary included in \cite{Harlow:2019yfa}, and a related application in the quantization of gauge theory in \cite{He:2020ifr}.

 \section{Symplectic 2-form on the space of solutions}
In this section, we review the formalism for constructing the covariant symplectic 2-form on the space of solutions of the equations of motion and show how this covariant form is related to the symplectic structure on the phase space. The major idea follows  \cite{woodhouse1997geometric, souriau1997structure}.
\subsection{The space of solutions as a manifold}
Given the action $I = \int_{t_i}^{t_f}L(x^i,v^i)dt$ of the dynamical space $S$ with $n$ degrees of freedom, the solutions to the equations of motion:

 \begin{equation*}
 \frac{d}{dt}(\frac{\partial L}{\partial v^i})- \frac{\partial L}{\partial x^i}=0 \eqno{(1)}
 \end{equation*}
 form the set of the space of motions $\mathcal{M}$. The manifold structure on $M$ is constructed by the diffeomorphism \cite{woodhouse1997geometric}:
 
 \begin{equation*}
			\tau_{s} : \mathcal{M} \rightarrow TX: q \mapsto (x^i|_{t=s}, v^i|_{t=s}), i = 1...n \eqno{(2)}
\end{equation*}
where $TX$ is the velocity phase space of the system $S$ at the moment $t=s$ \cite{woodhouse1997geometric}. The configuration space $X$ of $S$ is a $n-$dimensional smooth manifold, with coordinates $(x^1,...,x^n)$. The velocity phase space $TX$ is the tangent bundle of $X$, with $2n$ coordinates $(x^1,...,x^n, v^1,...,v^n)$. In $(2)$, $q \in \mathcal{M}$ is a solution to the equations of motion $(1)$, and  $(x^i|_{t=s}, v^i|_{t=s})$ are the values of $(x^i, v^i)$ at time $t = s$. To make the notation simpler, we will use $x^i_s$, $v^i_s$ to denote $x^i|_{t=s}, v^i|_{t=s}$ in the following. 

The meaning of (2) is that when a solution to the equations of motion $(1)$ can be uniquely determined by the condition $(x^i|_{t=s}=x^i_s, v^i|_{t=s}=v^i_s)$  \cite{woodhouse1997geometric}, the map $\tau_s$ gives an isomorphism, and the space of motions $\mathcal{M}$ becomes a $2n-$dimensional manifold by assigning each point of $\mathcal{M}$ the coordinates $(x^1_s,...x^n_s, v^1_s,...,v^n_s)$. In this article, we only considered models that satisfy this condition.

\subsection{The covariant sympletic 2-form}
The covariant sympletic 2-form $\omega$ is a closed non-degenerate 2-form on $\mathcal{M}$.  It can be constructed as follows:

Let $\delta q = (\delta x^i, \delta v^i)$ be a tangent vector that linearizes the equations of motion $(1)$. The action changes along the direction of $\delta q$ as \cite{woodhouse1997geometric}
\begin{equation*}
			\begin{aligned}
				dI(\delta q) &= \int_{t_i}^{t_f} \left(\frac{\partial L}{\partial x^i}\delta x^i + \frac{\partial L}{\partial v^i}\delta v^i \right) dt \\
				&= \frac{\partial L }{\partial v^a} \delta x^i|_{t=t_f} -  \frac{\partial L }{\partial v^a} \delta x^i|_{t=t_i}+ \int_{t_i}^{t_f}\left[\frac{\partial L}{\partial x^i}- \frac{d}{dt}(\frac{\partial L}{\partial v^i})\right]\delta x^i dt\\
				&= \theta_{t_f} (\delta q) - \theta_{t_i}(\delta q) + \int_{t_i}^{t_f}\left[\frac{\partial L}{\partial x^i}- \frac{d}{dt}\left(\frac{\partial L}{\partial v^i}\right)\right]\delta x^i dt
			\end{aligned} \eqno{(3)}
		\end{equation*}
where $\theta_t$ is the sympletic potential one-form on $\mathcal{M}$ defined by

\begin{equation*}
	\begin{aligned}
		\theta_{t} (\delta q) =  \frac{\partial L }{\partial v^a} \delta x^i|_{t} 
	\end{aligned} \eqno{(4)}
\end{equation*}
The last term in the third line of equation (3) vanishes, as each point on $\mathcal{M}$ is a solution to the equations of motion.
The covariant symplectic 2-form $\omega$ is constructed from the potential form $\theta_t$ by taking its exterior derivative:

\begin{equation*}
	\omega = d\theta_t \eqno{(5')}
\end{equation*} from which we have the covariant form $\omega$ acting on the vector fields as:

\begin{equation*}
 			\begin{aligned}
 				& \omega(\delta_1q, \delta_2q)
 				 				= \frac{\partial^2 L}{\partial x^a \partial v^b}(\delta_1x^a \delta_2x^b-\delta_2x^a \delta_1x^b)  + \frac{\partial^2 L}{\partial v^a \partial v^b}(\delta_1v^a \delta_2x^b - \delta_2v^a \delta_1x^b ) 
 		\end{aligned} \eqno{(5)}
 		\end{equation*}
 	where the lower index $1, 2$ label out the different paths that define the tangent vector fields. There are several points that we should note:
 		\begin{itemize}
 			\item [1)]  The 2-form $\omega$ constructed from the potential 1-form $(4)$ is generally not non-degenerate. A degenerate form $\omega$ is called a presymplectic form \cite{woodhouse1997geometric}. This happens, for example, when there are gauge symmetries in the system and the direction of the degeneracy of the presymplectic form $\omega$ is the gauge direction, e.g. in \cite{Torre:2002xt}. To make the presymplectic form $\omega$ nondegenerate, a known reduction process can be applied \cite{doi:10.1063/1.528801}. In the models we studied here, the covariant form $\omega$ by $(4)$ is nondegenerate in the first place, so the reduction process will not be discussed here.
 			\item [2)] The form $\omega = d\theta_t $ by $(5)$ is independent of the choice of  time $t$. This can be seen from equation $(5')$, which gives $\omega_{t_1} -\omega_{t_2}=d \theta_{t_1}-d \theta_{t_2} =d(d I) =0$ \cite{woodhouse1997geometric}. This covariance makes the solution space a covariant phase space which encodes all the information of the dynamical evolution of the system intrinsically \cite{ashtekar1991covariant}.
 			
 			\item [3)] Using the diffeomorphism $\tau_s$ in $(2)$, we have $(\tau_s)_{\star} \omega = {\omega_L}_s$ where ${\omega_L}_s$ is the symplectic 2-form of the usual phase space at time $t=s$. Given the Hamiltonian function $H$ of the system, Hamilton's equations are equivalent to requiring 
 		
 			\begin{equation*}
 				X \lrcorner \omega_L +dH= 0 \eqno (6)
 			\end{equation*}

 			 where $X$, called the Hamiltonian vector field, is the tangent vector field of the dynamical trajectory. The dynamical trajectory parameterized by the time $t$ induces a one-parameter  diffeomorphism group $\varphi_t$ acting on the phase space. The condition $(6)$ tells us that $\varphi_t$ is not only diffeomorphism but also symplectomorphism, as the symplectic form $\omega_L$ is Lie dragged along the dynamical trajectory \cite{ashtekar1991covariant}, which is consistent with the fact that the covariant form $\omega$ is independent of time. More details about this point will be discussed in Section $3.1$.
 			 
 			 \end{itemize}

\section{Quantum dynamics}
In this section we discuss how the classical physics is described in the language of covariant phase space, mainly based on the discussion in \cite{woodhouse1997geometric, souriau1997structure}, and  the quantization process for the covariant phase space, which will be applied to models in the later sections.
\subsection{Classical Physics}
In Section $2.1$ we saw that the covariant phase space becomes a manifold, given the diffeomorphism $(2)$, where the phase space $TX$ at time $t=s$ provides a set of coordinates of the manifold $\mathcal{M}$. In the last part of Section $2.2$ we noted that the push forward of the covariant form $(\tau_s)_{\star} \omega$ gives a symplectic structure ${\omega_L}_s$ on the phase space. Choosing two distinct moments $t=0$ and $t = s$ on the dynamical trajectory, the diffeomorphisms $\tau_0$ and $\tau_s$ give two different coordinate charts for the covariant phase space, as shown in Figure 1. 

\begin{figure}[htb]
\centering
\begin{center}
\[\xymatrix{
		 & & & &*+++=[o][F] {TQ|_{t=s}, {\omega_L}_s} \ar@{.>}@/^5pc/"3,5"^{(\tau_s \circ \tau_{0}^{-1})^{*} {\omega_L}_s = {\omega_L}_0} \ar@{.>}@/_5pc/"2,1"^{\omega_{\mathcal{M}} = \tau_s^{*} {\omega_L}_s} \\
		*++++=[o][F] {\mathcal{M}, \omega_{\mathcal{M}}}\ar@/^2pc/"1,5"^{\tau_{s}} \ar@/_2pc/"3,5"^{\tau_0}& & &\\
		 & & & &*+++=[o][F] {TQ|_{t=0}, {\omega_L}_0} \ar@/_1pc/"1,5"^{\varphi_s = \tau_{s} \circ \tau_{0}^{-1}} \ar@{.>}@/_1pc/"2,1"^{\tau_{0}^{-1}}
		} \]

Figure [1] 

\end{center}
\end{figure}
As we can see in Figure 1, the dynamical trajectory induces a one-parameter diffeomorphism group $ \lbrace \varphi_s \rbrace$ labeled by $s$. Note that the diffeomophism $\varphi_s = \tau_s \circ \tau_{0}^{-1}$ actually plays two roles here. On one hand, seen from the perspective of the usual classical physics,  $\varphi_s$ tells us how the system evolves from the moment $t=0$ to the moment $t=s$ along the dynamical trajectories as the orbits of the Hamiltonian vector field. On the other hand, using the language of the covariant phase space, we see that $\varphi_s = \tau_s \circ \tau_{0}^{-1}$ is also the transition function between two coordinate charts of the manifold $\mathcal{M}$ of the covariant phase space.

These two viewpoints are compatible with each other. As with the condition $(6)$, we know that the Lie derivative of the symplectic form $\omega_L$ vanishes along the dynamical trajectory, so

\begin{equation*}
	{\omega_L}_0 =(\varphi_s)^{*} {\omega_L}_s= (\tau_s \circ \tau_0^{-1} )^{*} {\omega_L}_s \eqno{(7)}
\end{equation*}
This is consistent with the requirement of the covariant form that

\begin{equation*}
	\omega_{\mathcal{M}}= \tau_{s}^{*} {\omega_L}_{s} =\tau_{0}^{*} {\omega_L}_{0} \eqno{(8)}
\end{equation*}
since from $(8)$  
\begin{equation*}
  (\varphi_s)^{*} {\omega_L}_s=(\tau_s \circ \tau_0^{-1} )^{*} {\omega_L}_s =(\tau_0^{-1})^{*}(\tau_s)^{*} {\omega_L}_s = (\tau_0^{-1})^{*} \omega_{\mathcal{M}} = {\omega_L}_0
\end{equation*}
which is exactly what we have in $(7)$.
The two roles played by $\varphi_s$ show that the evolution of the system with respect to time parameter $t$ in classical physics can also be described as the change of the coordinates of the manifold of the space of solution.

\subsection{Quantization}
When the space of solutions is uniquely determined by the set of values of solutions at some specific moment, these values can be considered as  coordinates on the space of solutions. By treating these coordinates as operators, we can find their eigenfunctions. These eigenfunctions form a basis of the Hilbert space which gives a representation of the Hilbert space. We quantize the covariant phase space in these Hilbert space representations. 
The major steps are as follows:
\begin{itemize}
	\item [1)] Starting from the Lagrangian, we find a covariant symplectic 2-form and  equations of motions by using the formulas of Section 2.1.
	\item [2)] We find a complete set of solutions to the equations of motion.  This set can be uniquely specified by a set of values of solutions at some specific moment. We use these values at a particular moment as coordinates on the space of solutions.
	\item [3)] Treating these coordinates as operators (``coordinate operators''), we find their eigenstates, which form a basis for the Hilbert space.  This gives a specific representation of the Hilbert space. The commutation relations between these operators are read from the symplectic covariant 2-form.
	\item [4)] We construct the Hamiltonian operator from the coordinate operators, and express the eigenstates of the Hamiltonian  in the representation obtained in $3)$. 
	\item [5)] Our coordinate operators are related to each other by equations of motion, with the time parameter $t$ as a label that distinguishes different sets of coordinates. The inner products between the eigenstates of different sets of coordinate operators are the kernels of the transformation among different representations.
	\item [6)] To explore the time evolution nature of the quantum states, we switch from one representation to another by using the kernels obtained in $5)$---this is simply performing the change of basis in Hilbert space---to see if there is an extra time dependent phase appearing in the new representation.
	\end{itemize}
It will be shown in Section 4 that the Hamiltonian operator and commutation relations remain invariant when they are constructed from different sets of the coordinate operators labeled by the time parameter.  In our non-relativistic model, we will see that an extra time dependent phase appears when we switch the basis, exactly matching the one that appears when the time evolution operator acts on stationary states in the usual quantization methods \cite{shankar1994principles}, while no such phase appears in our relativistic models.

\section{Some models}
In this section we quantize the covariant phase space of one non-relativistic model---harmonic oscillator---and three relativistic  cosmological models, using the quantization process of Section $3.2$.

\subsection{The Nonrelativistic Case: Harmonic Oscillator}
We start by quantizing the covariant phase space of the harmonic oscillator. By using the formalism in Section $2.2$ and solving the equations of motion, we have a set of complete solutions,

\begin{equation*}
	\begin{aligned}
		\begin{cases}
			 x_t= x_0cos(\omega t) + (v_{x_0}/ \omega m) sin(\omega t)   \\
	 		v_{x_t}= v_{x_0}cos(\omega t) - (x_0/ \omega m) sin(\omega t)
		\end{cases}
	\end{aligned} \eqno{(9)}
\end{equation*}
where $\omega$ is the frequency of the harmonic oscillator. Here $x_0 = x|_{t=0}, v_{x_0} = \dot x|_{t=0}$ are initial values of the solutions and $x_t = x|_{t}, v_{x_t} = \dot x|_{t}$ are values of the solutions at time $t$. A complete set of solutions can be specified by the initial values $(x_0, v_{x_0})$ or  by the values $(x_t, v_{x_t})$ at some other specific moment. These two sets  can be considered as two sets of coordinates of the manifold of the space of solutions, and the equations $(9)$ give the transition functions between these two coordinate charts, while also describing the dynamical trajectory of the system. The mapping of the coordinates $(x_0, v_{x_0})$ to the coordinates $(x_t, v_{x_t})$ is realized by the diffeomorphism $\varphi_t$ induced by the dynamical trajectory as discussed in Section $3.1$. To simplify the symbol, we use $v_t$ for $v_{x_t}$ in the following.

 The Poisson structure of these coordinates can be read from the covariant form by using $(5)$, which gives
 
 \begin{equation*}
	\begin{aligned}
	 \lbrace x_0, v_0 \rbrace = 1  \quad \quad \quad 
	 \lbrace x_s, v_s \rbrace = 1 
	\end{aligned} \eqno{(10')}
\end{equation*}
 As we can see, the Poisson structure is independent on the time parameter $t=s$. Treating these two sets of coordinates $(x_0, v_{x_0})$ and $(x_s, v_{x_s})$ as two sets of coordinate operators, we have the commutation relations

\begin{equation*}
\begin{aligned}
	[\hat x_0, \hat v_0]=i  \quad \quad \quad 
	[\hat x_s, \hat v_s]=i  
\end{aligned} \eqno{(10)}
\end{equation*}

We can construct the Hamiltonian operator from the coordinate operator $(\hat x_s, \hat v_{s})$ and their commutation relations $(10)$.  The eigenstates of the Hamiltonian operator in the $\hat x_s$-representation are
  \begin{equation*}
    	\begin{aligned}
    		\langle x_s | \hat  H | \psi_{ns} \rangle &= \langle x_s | \frac{\hat v_s^2}{2m} + \frac{1}{2}m \omega^2 \hat x_s^2 | \psi_{ns} \rangle \\
    	&= \left(- \frac{1}{2m} \frac{\partial^2 }{\partial x_s^2}  + \frac{1}{2}m \omega^2  x_s^2 \right) \psi_{ns}(x_s) = E_n \psi_{ns}(x_s)
    	\end{aligned} \eqno{(11)}
    \end{equation*}
 where $|x_s \rangle$ is an eigenstates of $\hat x_s$ with eigenvalue $x_s$. Here  $| \psi_{ns} \rangle $ are eigenstates of the Hamiltonian operator, and $\psi_s(x_{ns}) \equiv \langle x_s| \psi_{ns} \rangle$ are these eigenstates in the $\hat x_s$-representation. Solving $(11)$ we have
    
      \begin{equation*}
    	\psi_{ns}(x_s) = \frac{1}{\sqrt{2^n n!}} (\frac{m \omega}{\pi})^{\frac{1}{2}} e^{-\frac{m \omega x_s^2}{2}} H_n(\sqrt{m \omega} x_s )  \eqno{(12)}
    \end{equation*} where $H_n$ is the Hermite functions. Similarly, we have  eigenstates $| \psi_{0n} \rangle$ of the Hamiltonian in $\hat x_0$-representation:     
  \begin{equation*}
    	\begin{aligned}
    		\langle x_0 | \hat  H | \psi_{0n} \rangle &= \langle x_0 | \frac{\hat v_0^2}{2m} + \frac{1}{2}m \omega^2 \hat x_0^2 | \psi_{0n} \rangle \\
    	&= (- \frac{1}{2m} \frac{\partial^2 }{\partial x_0^2}  + \frac{1}{2}m \omega^2  x_0^2 ) \psi_{0n}(x_0) = E_n \psi_{0n}(x_0)
    	\end{aligned} \eqno{(13)}
    \end{equation*}
    from which      
     \begin{equation*}
    	\psi_{0n}(x_0) = \frac{1}{\sqrt{2^n n!}} (\frac{m \omega}{\pi})^{\frac{1}{2}} e^{-\frac{m \omega x_0^2}{2}} H_n(\sqrt{m \omega} x_0 ) \eqno{(14)}
    \end{equation*} 
From $(10)$ and $(9)$,  the kernel of the transformation between these representations satisfies
    
    \begin{equation*}
    	\begin{aligned}
    		\begin{cases}
    				(x_0cos(\omega s) - i \mu \frac{\partial}{\partial x_0 } sin(\omega s))  \langle x_0| x_s \rangle = x_s  \langle x_0|  x_s \rangle    \\
		(x_scos(\omega s) + i \mu \frac{\partial}{\partial x_0 } sin(\omega s))  \langle x_s| x_0 \rangle = x_0  \langle x_s|  x_0  \rangle   		\end{cases}
    	\end{aligned}.  \eqno{(15)} 
    \end{equation*}
   with $\mu = (m\omega)^{-1}$. The solution is
      \begin{equation*}
   		\langle x_0 | x_s \rangle = exp\left\{i\left(-\frac{1}{2 \mu \tan(\omega s)}(x_0^2 + x_s^2)+\frac{x_s x_0}{\mu sin(\omega s)} \right) \right\} \eqno{(16)}
   \end{equation*}
   Note that the mathematical process of using equations $(15)$ to get $(16)$ is not new. It appeared in \cite{1937PNAS...23..158C}, where $(16)$ is called as the Mehler kernel, but with different interpretation coming
  purely from the mathematical aspect. The Mehler kernel has also been used in the path integral, but again by a different method and with a different physical interpretation. 
   
   To recover the time evolution of the quantum states, we switch the $\hat x_s$-representation to the $\hat x_0$-representation by using the kernel $(16)$:
   \begin{equation*}
   	\begin{aligned}
   			\psi_{sn}(x_0) & = \int \langle x_0| x_s \rangle \langle x_s | \psi_{sn} \rangle dx_s\\ 
   		 &= \int  \mathcal{C} e^{i(-\frac{1}{2 \mu \tan(\omega s)}(x_0^2 + x_s^2)+\frac{x_s x_0}{\mu sin(\omega s)} ) }e^{-\frac{m \omega x_s^2}{2}} H_n(\sqrt{m \omega} x ) dx_s\\ 
   		 &= e^{- in \omega (-s)}  \frac{1}{\sqrt{2^n n!}} (\frac{m \omega}{\pi})^{\frac{1}{2}} e^{-\frac{m \omega x_0^2}{2}} H_n(\sqrt{m \omega} x_0 )\\
   		 &=  e^{- in \omega (-s)} \psi_{0n}(x_0)
   	\end{aligned} \eqno{(17)}
   \end{equation*}where the details of this integral can be found in the Appendix. We see that in a non-relativistic model such as the harmonic oscillator, when we transform from the $\hat x_s$-representation to the $\hat x_0$-representation, a time dependent phase $e^{-inw(-s)}$ appears. This extra  phase is exactly the one produced by the time evolution operator $e^{-i \mathcal{\hat H} t}$ acting on the stationary state with eigenvalue $E_n = n \hbar \omega$  when the usual quantization method is applied. We also note that the quantum states in each representation have the same expression as states in the usual position space representation \cite{shankar1994principles}. This is because the commutation relations in both cases have the same form. The difference is that in the usual quantization method  the quantization is performed on the phase space at a specific time, while here it is the covariant phase space that is quantized.

\subsection{Relativistic Cases: Cosmological models}
We now quantize the covariant phase space of three cosmological models as examples of relativistic models. The models considered below were chosen from the literature
   more or less at random as typical examples of cosmological
   models that can be exactly quantized by other means. Note that each model involves a lapse function, which appears in the Hamiltonian only as a Lagrange multiplier. By a gauge choice, this function can be quantity to be any nonnegative function; in the models investigated here, the particular choices are made for calculational convenience. We will see that in each case, no extra time dependent phase appears when we switch representations, indicating that the quantum states of these relativistic models do not evolve along a particular time $t$ direction.
\subsubsection{Example 1: FRW Universe coupled with dilaton field}
Our first relativistic model is a spatially flat Friedmann-Roberston-Walker (FRW) Universe coupled with a dilaton field. The metric of this model that describes the spacetime curvature is \cite{PhysRevD.77.044023}:

\begin{equation*}
	\begin{aligned}
		ds^2= - \frac{N^2(t)}{a^2(t)}dt^2+a^2(t) d \Omega_3^2 
	\end{aligned}
\end{equation*}
of which $a^2(t)$ is the scale factor, $N^2(t)/a^2(t)$ is the lapse function and $d \Omega_3^2$ is the standard metric on $S^3$. The effect of the dilaton field on this model is introduced by the Einstein-Hilbert action \cite{PhysRevD.77.044023}:

\begin{equation*}
	\begin{aligned}
		\mathcal{S} = \int d^4x \sqrt{-g} ( \mathcal{R} - \frac{1}{2} \partial_{\mu}\partial^{\mu} \phi - V(\phi))
	\end{aligned}
\end{equation*}
where $\mathcal{R}$ is the scalar curvature of the spacetime,  $\phi$ is the dilton field and $V(\phi ) = (V_0/2)e^{- \alpha \phi}$ is the potential of the dilton field with $\alpha$ and $V_0$ two positive constants. 
Using the metric and the Einstein-Hilber action, we get the Lagrangian of this model as \cite{PhysRevD.77.044023}

\begin{equation*}
	\begin{aligned}
		\mathcal L= \frac{1}{N}\left(- \frac{1}{2}a^2 \dot a^2 + \frac{1}{2} a^4 \dot \phi ^2 \right) -N a^2 V(\phi )
	\end{aligned} \eqno{(18)}
\end{equation*}
To simply the Lagrangian, we do the change of variables as follows:
 \begin{equation*}
 	\begin{aligned}
 		\left\{
\begin{array}{c}
	x=(a^2/2) \cosh(\alpha\phi)\\ 
	y=(a^2/2) \sinh(\alpha\phi)

\end{array}\right.
 	\end{aligned}  
 \end{equation*}
 which gives \cite{PhysRevD.77.044023}:
 \begin{equation*}
 	\begin{aligned}
 		\mathcal L= \frac{1}{2N}(\dot y^2-\dot x^2) - NV_0(x-y)
 	\end{aligned} \eqno{(19)}
 \end{equation*}
 Setting $N =1$ and using the formalism $(3)$ in Section $2.2$, we have 
\begin{equation*}
	\begin{aligned}
	d	\mathcal L (\delta q)
			= \underbrace{\frac{d}{dt}\left(\frac{dy}{dt}\delta y-\frac{dx}{dt}\delta x\right)}_{\text{boundary term}} 
		 + \underbrace{\left(\frac{d^2x}{dt^2}-V_0\right)\delta x + \left(-\frac{d^2 y}{dt^2}+V_0\right)\delta y}_{\text{bulk term}}
	\end{aligned} \eqno{(20)}
\end{equation*}
From the boundary term we get the potential 1-form

\begin{equation*}
	\begin{aligned}
		\theta_t(\delta q) = \frac{dy_t}{dt}\delta y_t-\frac{dx_t}{dt}\delta x_t =v_{y_t} \delta y_t - v_{x_t} \delta x_t
	\end{aligned} \eqno{(21)}
\end{equation*} with $dx_t/dt \equiv v_{x_t}, dy_t/dt \equiv v_{y_t}$, 
and from the bulk term, we obtain the equations of motion

\begin{equation*}
	\begin{aligned}
		\begin{cases}
	\ddot x- V_0 = 0    \\
	\ddot y- V_0= 0
	\end{cases}	
	\end{aligned} 
\end{equation*}
A set of complete solutions is

\begin{equation*}
	\begin{aligned}
		\begin{cases}
	 x_t= \frac{1}{2} V_0 t^2 + v_{x_0} t + x_0  \\

	 y_t= \frac{1}{2} V_0 t^2 + v_{y_0} t + y_0  
	\end{cases} 
	\end{aligned} \eqno{(22)}
\end{equation*}
where $x_0 = x|_{t=0}, v_{x_0} = \dot x|_{t=0}$ and $y_0 = y|_{t=0}, v_{y_0} = \dot y|_{t=0}$ are initial values. Similarily, $x_s = x|_{t=s}, v_{x_s} = \dot x|_{t=s}$ and $y_s = y|_{t=s}, v_{y_s} = \dot y|_{t=s}$ are values at $t=s$. As discussed in previous sections, a set of complete solutions can be specified by the initial values $(x_0, v_{x_0}, y_0, v_{y_0})$ or  by the values at some specific moment  $(x_s, v_{x_s}, y_s, v_{y_s})$, giving two sets of the coordinates of the manifold of the space of motions. The equations $(22)$ give the transition functions between these two sets of coordinates. Additionally, by using $(22)$, we have
\begin{equation*}
\begin{aligned}
	\begin{cases}
	 v_{x_t} = V_0 t+ v_{x_0}\\

	v_{y_t} = V_0 t+ v_{y_0}	\end{cases}	
	\end{aligned}
	\quad \quad \quad 
	\begin{gathered}
	\begin{cases}
	 \delta x_t = t \delta v_{x_0}  + \delta x_0\\

	 \delta y_t = t \delta v_{y_0}  + \delta y_0 \\
	\end{cases}
	\end{gathered} \eqno{(23)}
\end{equation*}
Inserting $(23)$ into $(21)$, we get the potential 1-form

\begin{equation*}
	\begin{aligned}
		\theta_t (\delta q) &= (V_0t+ v_{y_0})(t \delta v_{y_0}+ \delta y_0)- (V_0t+ v_{x_0})(t \delta v_{x_0} + \delta x_0)\\
	\end{aligned} \eqno{(24)}
\end{equation*}
which does now depend on the time parameter $t$. Using $(5)$, we find the covariant symplectic 2-form
\begin{equation*}
	\begin{aligned}
		 \omega(\delta_1q, \delta_2q) 
 				 				 &=  (\delta_1v_{y_t} \delta_2 y_t - \delta_2v_{y_t} \delta_1 y_t)  - (\delta_1v_{x_t} \delta_2 x_t - \delta_2v_{x_t} \delta_1 x_t)\\
 				 				 &= (\delta_1v_{y_0} \delta_2 y_0 - \delta_2v_{y_0} \delta_1 y_0)  - (\delta_1 v_{x_0} \delta_2 x_0 - \delta_2v_{x_0} \delta_1 x_0)
\end{aligned} \eqno{(25)}
\end{equation*}
which is independent of $t$. We obtain the Poisson structure from $(25)$:
\begin{equation*}
\begin{aligned}
	\begin{cases}
	 \lbrace v_{x_0} , x_0 \rbrace=-1 \\

	\lbrace v_{y_0} , y_0\rbrace= 1\end{cases}	
	\end{aligned}
	\quad \quad \quad 
	\begin{gathered}
	\begin{cases}
	 \lbrace v_{x_t}, x_t \rbrace=-1\\

	 \lbrace v_{y_t}, y_t\rbrace= 1 \\
	\end{cases}
	\end{gathered}\eqno{(26')}
\end{equation*} Treating the set of coordinates $(x_t, v_{x_t}, y_t, v_{y_t})$ as operators, we have the commutation relations:

\begin{equation*}
\begin{aligned}
	\begin{cases}
	[ \hat v_{x_0} , \hat x_0 ]=-i \\

	[ \hat v_{y_0} , \hat y_0]= i\end{cases}	
	\end{aligned}
	\quad \quad \quad 
	\begin{gathered}
	\begin{cases}
	[ \hat v_{x_t} , \hat x_t ] =-i\\

	[ \hat v_{y_t} , \hat y_t ] = i \\
	\end{cases}
	\end{gathered}\eqno{(26)}
\end{equation*} The Hamiltonian constraint from $(19)$ is constructed as follows. The conjugate momenta are:

\begin{equation*}
	 v_x=\frac{\partial \mathcal L}{\partial\dot x} =- \frac{1}{N} \dot x  \quad \quad \quad
	 v_y=\frac{\partial \mathcal L}{\partial\dot y} = \frac{1}{N} \dot y 
\end{equation*}. The Hamiltonian  at moment $t$ is

\begin{equation*}
	\begin{aligned}
		\mathcal H|_t = N(- \frac{1}{2}(v_{x_t}^2-v_{y_t}^2)+V_0(x_t-y_t) )
	\end{aligned} \eqno{(27)}
\end{equation*} which turns out to be a constraint and should vanish, so the lapse function $N$ can be omitted in the following \cite{PhysRevD.77.044023}. Inserting $(22)$ into $(27)$, we have


\begin{equation*}
	\begin{aligned}
		\mathcal H|_t & =- \frac{1}{2}(v_{x_t}^2-v_{y_t}^2)+V_0(x_t-y_t)\\
				&=\frac{1}{2}(v_{y_0}^2-v_{x_0}^2)+V_0(x_0-y_0) = \mathcal H|_0
	\end{aligned}
\end{equation*}
which shows that the Hamiltonian is also independent of $t$. 

Now using the coordinate operators $(\hat x_t, \hat v_{x_t}, \hat y_t, \hat v_{y_t})$, the Hamiltonian can also be constructed as an operator and the eigenstates of the Hamiltonian can be expressed in the $(\hat x_s, \hat y_s)$-representation. Using $(26)$ and the Hamiltonian constraint $(27)$ at $t=s$, we have

\begin{equation*}
	\begin{aligned}
		\langle x_s, y_s| \hat{\mathcal{H}} | \psi \rangle &= \langle x_s, y_s|   \frac{1}{2}(\hat v_{y_s}^2-\hat v_{x_s}^2)+V_0(\hat x_s- \hat y_s) | \psi \rangle \\
		&= \frac{1}{2}\left[\frac{\partial^2}{\partial x_s^2}-\frac{\partial^2}{\partial y_s^2} + 2V_0(x_s-y_s)\right] \psi_s (x_s,y_s)=0
	\end{aligned} \eqno{(28)}
\end{equation*}
where $|x_t \rangle, |y_t \rangle$ are eigenstates of the coordinate operators $\hat x_t, \hat y_t$ and $| \psi \rangle $ is the eigenstate of the Hamiltonian operator with $\psi_s(x_s,y_s) \equiv \langle x_s, y_s | \psi \rangle $. To solve $(28)$, we follow the procedure in \cite{PhysRevD.77.044023}, obtaining solutions  similar to those in that reference:
\begin{equation*}
	\begin{aligned}
		\psi_s(x_s,y_s) = \psi_{s1}(x_s) \psi_{s2}(y_s)=Ai\left(\frac{\eta -2V_0 x_s}{(2V_0)^{2/3}}\right)Ai\left(\frac{\eta -2V_0 y_s}{(2V_0)^{2/3}}\right) 
	\end{aligned} \eqno{(29)}
\end{equation*}
where $\eta$ is a separation constant and $Ai$ is the Airy function. 

Similarly, in the coordinate ($\hat v_{x_0}, \hat v_{y_0}$) space, we have
\begin{equation*}
	\begin{aligned}
		\langle v_{x_0}, v_{y_0}| \hat{\mathcal{H}} | \psi \rangle = [\frac{1}{2}(v_{y_0}^2-v_{x_0}^2)+V_0(i\frac{\partial}{\partial v_{x_0}} + i\frac{\partial}{\partial v_{y_0}})] \psi_{v_0} (v_{x_0}, v_{y_0}) =0
	\end{aligned} \eqno{(30)}
\end{equation*}
where $|v_{x_0} \rangle, |v_{y_0} \rangle$ are eigenstates of $\hat v_{x_0}, \hat v_{y_0}$ and $| \psi \rangle $ is the eigenstate of the Hamiltonian operator with $\psi_{v_0}(v_{x_0},v_{y_0}) \equiv \langle v_{x_0}, v_{y_0} | \psi \rangle $. Solving $(30)$ we have

\begin{equation*}
	\begin{aligned}
		\psi_{v_0} (v_{x_0},v_{y_0}) &= \psi_1(v_{x_0}) \psi_2(v_{y_0})\\
		&= e^{i(-\frac{1}{6V_0} v_{x_0}^3 -\frac{1}{2V_0} \eta v_{x_0})}  e^{i(\frac{1}{6V_0} v_{y_0}^3 + \frac{1}{2V_0} \eta v_{y_0})} 
	\end{aligned} \eqno{(31)}
\end{equation*}
with $\eta$ the separation constant.

 As $\psi_s(x_s,y_s) \equiv \langle x_s, y_s | \psi \rangle $ and $\psi_{v_0}(v_{x_0},v_{y_0}) \equiv \langle v_{x_0}, v_{y_0} | \psi \rangle $ are both the eigenstates of the Hamiltonian operator, but expressed in different representations,\footnote{We can also express $ | \psi \rangle$ in the $(\hat x_0, \hat y_0)$-representation; the relation between $(\hat x_0, \hat y_0)$-representation and $( \hat v_{x_0}, \hat v_{y_0})$-representation is similar to that between the ordinary position and momentum representations.} it is interesting to see how they transform from one to the other. We first note that $(22)$ relates the coordinates operators that are labeled by the time parameter as
 
 \begin{equation*}
\begin{aligned}
	\end{aligned}
	\quad \quad \quad 
	\begin{gathered}
	\begin{cases}
	 \hat x_s=\frac{1}{2} V_0 s^2 + \hat v_{x_0} s + x_0 \\

	 \hat y_s = \frac{1}{2} V_0 s^2 + \hat v_{y_0} s + y_0  
	\end{cases} 
	\end{gathered}\eqno{(32)}
\end{equation*}
Combining $(32)$ with the commutation relations $(26)$, we have
\begin{equation*}
\begin{aligned}
	\end{aligned}
	\quad \quad \quad 
	\begin{gathered}
	\begin{cases}
 \langle x_0 | \hat x_s  | x_s \rangle =  (\frac{1}{2}V_0s^2 -i \frac{\partial }{\partial x_0} s+ x_0)  \langle x_0| x_s  \rangle =x_s \langle x_0 | x_s \rangle \\
 \\
	  \langle x_s | \hat x_0  | x_0 \rangle =  (\frac{1}{2}V_0s^2 +i \frac{\partial }{\partial x_s} s + x_s )\langle x_s | x_0  \rangle =x_0 \langle x_s | x_0 \rangle 
	\end{cases} 
	\end{gathered}\eqno{(33)}
\end{equation*}

\begin{equation*}
\begin{aligned}
	\end{aligned}
	\quad \quad \quad 
	\begin{gathered}
	\begin{cases}
 \langle y_0 | \hat y_s  | y_s \rangle =  (\frac{1}{2}V_0s^2 +i \frac{\partial }{\partial y_0} s+ y_0)  \langle y_0| y_s  \rangle =y_s \langle y_0 | y_s \rangle \\
 \\
	  \langle y_s | \hat y_0  | y_0 \rangle =  (\frac{1}{2}V_0s^2 -i \frac{\partial }{\partial y_s}s + y_s )\langle y_s | y_0  \rangle =y_0 \langle y_s | y_0 \rangle 
	\end{cases} 
	\end{gathered}\eqno{(33')}
\end{equation*}
Solving $(33)$ and $(33')$ we obtain the kernel that needed to perform the transformation between different representations:

\begin{equation*}
\begin{cases}
\begin{aligned}
	\langle x_0| x_s  \rangle &= e^{i(-\frac{1}{2s} (x_0^2 + x_s^2) -\frac{1}{2}V_0 s(x_0 + x_s) + \frac{1}{s}x_0x_s )}\\
	 \langle y_0| y_s  \rangle &= e^{i(\frac{1}{2s} (y_0^2 + y_s^2) +\frac{1}{2}V_0 s(y_0 + y_s) - \frac{1}{s}y_0y_s )}
\end{aligned} 
\end{cases} \eqno{(34)}
\end{equation*}
The commutation relations $(26)$ also imply

\begin{equation*}
\begin{cases}
\begin{aligned}
	 \left \langle v_{x_0}| \hat x_0|x_0 \right \rangle &=\left \langle v_{x_0}| i \frac{\partial}{\partial v_{x_0} } |x_0 \right \rangle = x_0 \left \langle v_{x_0}| x_0 \right \rangle \\
	  \left \langle v_{y_0}| \hat y_0|y_0 \right \rangle &=\left \langle v_{y_0}|- i \frac{\partial}{\partial v_{y_0} } |y_0 \right \rangle = y_0 \left \langle v_{y_0}| y_0 \right \rangle
\end{aligned}
\end{cases} \eqno{(35)}
\end{equation*} from which 

\begin{equation*}
\begin{cases}
\begin{aligned}
	  \langle v_{x_0}| x_0 \rangle &= e^{-iv_{x_0}x_0} \\
	  
	  \langle v_{y_0}| y_0 \rangle &= e^{iv_{y_0}y_0} \\
\end{aligned}
\end{cases} \eqno{(36)}
\end{equation*}
Combining $(34)$ with $(36)$ we have

\begin{equation*}
\begin{cases} 
\begin{aligned}
	   \langle v_{x_0}| x_s \rangle &= \int \langle v_{x_0}| x_0 \rangle \langle x_0| x_s \rangle dx_0  \\
	  
	  \langle v_{y_0}| y_0 \rangle &=  \int \langle v_{y_0}| y_0 \rangle \langle y_0| y_s \rangle dy_0
\end{aligned}
\end{cases} \eqno{(37)}
\end{equation*}
Solving $(37)$ we get the expression of kernel as

\begin{equation*}
\begin{cases}
\begin{aligned}
	   \langle v_{x_0}| x_s \rangle & = e^{i(\frac{1}{2}v_{x_0}^2s + v_{x_0}(\frac{1}{2}V_0s^2 - x_s) + \frac{1}{8}V_0^2s^3 - sV_0x_s )} \\
	  
	  \langle v_{y_0}| y_0 \rangle &=  e^{-i(\frac{1}{2}v_{y_0}^2s + v_{y_0}(\frac{1}{2}V_0s^2 - y_s) + \frac{1}{8}V_0^2s^3 -s V_0y_s )}
\end{aligned}
\end{cases} \eqno{(38)}
\end{equation*}
With $(38)$, we can transform eigenstates $(29)$ of the Hamiltonian in ($\hat x_s, \hat y_s$)-representation to those in ($\hat v_{x_0}, \hat v_{y_0}$)-representation:
\begin{equation*}
\begin{aligned}
	\psi_{s1} (v_{x_0}) &=\int \langle v_{x_0} | x_s \rangle \langle x_s | \psi_{s1} \rangle dx_s 
	= \mathcal{C}e^{i(-\frac{1}{6V_0}v_{x_0}^3 -\frac{\eta v_{x_0}}{2V_0})}e^{i(-\frac{1}{24}V_0^2s^3- \frac{s}{2} \eta )} 
\end{aligned} \eqno{(39)}
\end{equation*} The details of the integral are in Appendix B.
Similarly,
\begin{equation*}
\begin{aligned}
\psi_{s2}(v_{y_0}) &= \int \langle v_{y_0} | y_s \rangle \langle y_s | \psi_{s2} \rangle dy_s =\mathcal{C} e^{i(\frac{1}{6V_0}v_{y_0}^3 +\frac{\eta v_{x_0}}{2V_0})}e^{i(\frac{1}{24}V_0^2s^3+ \frac{s}{2} \eta )}
	\end{aligned} \eqno{(40)}
\end{equation*}
where $\mathcal{C}$ is a constant fixed by the normalization condition. Combining $\psi_{s1}(v_{x_0})$ and $\psi_{s2}(v_{y_0})$, the eigenstates $\psi_s(x_s, y_s)$ in the coordinate-$(\hat v_{x_0}, \hat v_{y_0})$ representation can be transformed to those in the coordinate-$(\hat v_{x_0}, \hat v_{y_0})$ representation as
\begin{equation*}
	\begin{aligned}
		\psi_s(v_{x_0},v_{y_0}) 
		&= \psi_{s1}(v_{x_0}) \psi_{s2}(v_{y_0}) 
				= e^{i(-\frac{1}{6V_0} v_{x_0}^3 -\frac{1}{2V_0} \eta v_{x_0})}  e^{i(\frac{1}{6V_0} v_{y_0}^3 + \frac{1}{2V_0} \eta v_{y_0})} \\
		&= \psi_{v_0} (v_{x_0}, v_{y_0})
	\end{aligned} \eqno{(41)}
\end{equation*}
(see Appendix B for details). 
In this example, we also see two features:

 $1)$ As seen in $(28)$, the quantum states have the same expressions in the covariant phase space representations as those in the position representation obtained by the usual quantization method of \cite{PhysRevD.77.044023}. This is reasonable, as the symplectic 2-form in the phase space at a specific moment has the same form as the covariant symplectic 2-form in the covariant phase space. 
 
$2)$ From $(41)$ we see that, in contrast to the non-relativistic model of previous section, no extra time dependent phase appears in the quantum states when we switch from one representation to another.  This indicates that in our relativistic model, the quantum states do not evolve along a particular coordinate time $t$ direction. This is consistent with the result of other quantization methods \cite{PhysRevD.77.044023}. The same conclusions will also be drawn in other relativistic models below.

\subsubsection{Example 2: FRW Universe with cosmological constant}
The second relativistic model we study is a spatially flat Friedmann–Robertson–Walker (FRW) Universe with a cosmological constant $\Lambda$ with the metric \cite{Bina:2007wj}
\begin{equation*}
	\begin{aligned}
		ds^2 = -N^2(t) dt^2 + a^2(t)(dr^2 + r^2 d\Omega^2)
	\end{aligned}
\end{equation*} where $N(t)$ is the lapse function, $a(t)$ is the scale factor of the Universe and $d\Omega^2$ is the standard metric on $S^2$. The Einstein-Hilbert action of this model is \cite{Bina:2007wj}

\begin{equation*}
	\begin{aligned}
		\mathcal{S} = \int d^4x \sqrt{-g}(\mathcal{R} - \frac{1}{2} \partial_{\mu}\partial^{\mu} \phi - \Lambda)
	\end{aligned}
\end{equation*}where $\mathcal{R}$ is the scalar curvature, $\Lambda$ is the cosmological constant and $\phi$ is a free scalar field. Taking the metric to the action we get the Lagrangian of this model as \cite{Bina:2007wj}:


\begin{equation*}
	\begin{aligned}
		\mathcal{L} = \frac{1}{N}(-3a\dot a^2+\frac{1}{2} \dot\phi^2 a^3)- N a^3 \Lambda 
	\end{aligned} \eqno{(42)}
\end{equation*} Here we consider a negative cosmological constant $\Lambda < 0$ and set $N = 1$. By changing variables \cite{Bina:2007wj},

\begin{equation*}
	\begin{aligned}
	\begin{cases}
		x=a^\frac{3}{2}\cosh(\alpha\phi)\\ 
		y=a^\frac{3}{2}\sinh(\alpha\phi)
	\end{cases}		
		\end{aligned} 
\end{equation*}
with $\alpha ^2 = \frac{3}{8}$, the Lagrangian $(42)$ is simplified to \cite{Bina:2007wj}:
\begin{equation*}
	\begin{aligned}
		\mathcal{L} = ( -\frac{4}{3})(\dot x^2-\dot y^2 - \epsilon^2 (x^2-y^2)) 
	\end{aligned} \eqno{(43)}
\end{equation*} with $\epsilon^2 =-2 \alpha^2\Lambda > 0$, so $\epsilon$ is real. Using $(3)$ and the Lagrangian $(42)$, we can read equations of motion from the bulk term in $(3)$, with solutions

\begin{equation*}
\begin{aligned}
	\begin{cases}
	\ddot x +\epsilon ^2x= 0    \\
	\ddot y +\epsilon ^2y= 0
	\end{cases}	
	\end{aligned}
	\quad \quad \quad 
	\begin{gathered}
	\begin{cases}
	 x_t= x_0\cos(\epsilon t) + v_{x_0}\sin(\epsilon t)  \\

	 y_t= y_0\cos(\epsilon t) + v_{y_0}\sin(\epsilon t) 
	\end{cases}
	\end{gathered}\eqno{(44)}
\end{equation*}
The potential 1-form from $(3)$ is

\begin{equation*}
	\theta_t(\delta q) = (-8/3) (v_{x_t} \delta x_t- v_{y_t} \delta y_t)
\end{equation*} from which we have the symplectic covariant 2-form
\begin{equation*}
	\begin{aligned}
	\omega(\delta_1 q, \delta_2 q ) &=	(-8/3) \epsilon [(\delta_1 v_{x_t} \delta_2 x_t -\delta_2 v_{x_t} \delta_1 x_t)- (\delta_1 v_{y_t} \delta_2 y_t- \delta_2 v_{y_t} \delta_1 y_t) ]\\
		&= (-8/3)\epsilon [(\delta_1 v_{x_0} \delta_2 x_0 -\delta_2 v_{x_0} \delta_1 x_0)- (\delta_1 v_{y_0} \delta_2 y_0- \delta_2 v_{y_0} \delta_1 y_0) ]
		\end{aligned} \eqno{(45)}
\end{equation*} which shows that the covariant symplectic 2-form does not depend on the time parameter $t$. We have the Poisson structure from $(45)$

\begin{equation*}
	\begin{aligned}
		\begin{cases}
			\lbrace x_0, v_{x_0} \rbrace =  \mu \\
			\lbrace y_0, v_{y_0} \rbrace = -\mu
			
		\end{cases}
	\end{aligned}
	\begin{gathered}
	\begin{cases}
	\lbrace x_s, v_{x_s} \rbrace =  \mu \\
	\lbrace y_s, v_{y_s} \rbrace = - \mu
	\end{cases}
	\end{gathered} 
\end{equation*} 
with $\mu = \frac{8}{3} \epsilon$. Treating the set of coordinates $(x_s, v_{x_s}, y_s, v_{y_s})$ as operators, we have commutation relations

\begin{equation*}
	\begin{aligned}
		\begin{cases}
			[\hat x_0, \hat v_{x_0}] = i \mu \\
			[\hat y_0, \hat v_{y_0}] = -i \mu
		\end{cases}
	\end{aligned}
	\begin{gathered}
	\begin{cases}
	[\hat x_s, \hat v_{x_s}] = i \mu \\
	[\hat y_s, \hat v_{y_s}] = -i \mu
	\end{cases}
	\end{gathered} \eqno{(46)}
\end{equation*}

The Hamiltonian constraint from $(42)$ is
\begin{equation*}
	\begin{aligned}
		\mathcal{H} &= \dot x \frac{\partial \mathcal{L}}{\partial \dot x} + \dot y \frac{\partial \mathcal{L}}{\partial \dot y}  - \mathcal{L} =(-4/3)(\dot x ^2 -\dot y^2 + \epsilon ^2(x^2-y^2)) \\
		&=(-4/3)(v_x ^2 -v_y ^2 + \epsilon ^2(x^2-y^2))
	\end{aligned} \eqno{(47)}
\end{equation*}
Inserting $(44)$ into $(47)$ we find that the Hamiltonian constraint $\mathcal{H}$ is independent of the time parameter $t$. In $(\hat x_s, \hat y_s)$-representation with commutation relations $(46)$ we have 
\begin{equation*}
	\begin{aligned}
		\langle x_s, y_s| \hat{\mathcal{H}} | \psi \rangle &= \langle x_0, y_0|   (-4/3)((\hat v_{x_s}^2-\hat v_{y_s}^2)+\epsilon^2 (\hat x_s ^2- \hat y_s ^2)) | \psi \rangle \\
		&= (- \frac{4}{3})\left[- \mu^2 \frac{\partial^2}{\partial x_s^2} + \mu^2 \frac{\partial^2}{\partial y_s^2}+ \epsilon^2(x_s^2 - y_s^2) \right]\psi_{s} (x_s, y_s) =0
	\end{aligned} \eqno{(48)}
\end{equation*}
which gives the eigenstates of the Hamiltonian in $(\hat x_s, \hat y_s)$-representation:
\begin{equation*}
	\begin{aligned}
		\psi_s(x_s, y_s) = \psi_{s1}(x_s) \psi_{s2}(y_s)
					   = \mathcal{C} e^{-\frac{1}{2\eta} x_s^2}e^{-\frac{1}{2\eta} y_s^2}H_n(\frac{1}{\sqrt{\eta}} x_s)H_n(\frac{1}{\sqrt{\eta}} y_s)	
	\end{aligned} \eqno{(49)}
\end{equation*}
where $\mathcal{C}$ is a constant, $\eta = \frac{3}{8} $, and $H_n$ is the Hermite function. Similarly we find the eigenstates of the Hamiltonian in $(\hat x_0, \hat y_0)$-representation:
\begin{equation*}
	\begin{aligned}
		\psi_0(x_0, y_0) = \psi_{s1}(x_0) \psi_{s2}(y_0)
					   = \mathcal{C} e^{-\frac{1}{2\eta} x_s^2}e^{-\frac{1}{2\eta} y_s^2}H_n(\frac{1}{\sqrt{\eta}} x_0)H_n(\frac{1}{\sqrt{\eta}} y_0)	
	\end{aligned} \eqno{(50)}
\end{equation*}
We can transform the quantum states from $(\hat x_s, \hat y_s)$-representation $(49)$ to those in the $(\hat x_0, \hat y_0)$-representation $(50)$ to see if a time dependent phase appears in the quantum states. 

 Let $|x_s \rangle, |y_s \rangle$ and $|x_0 \rangle, |y_0 \rangle$ be the eigenstates of coordinate operators $\hat x_s, \hat y_s$ and $\hat x_0 , \hat y_0$. We have:

\begin{equation*}
	\begin{aligned}
		\begin{cases}
			\langle x_0|\hat x_s |x_s \rangle = x_s  \langle x_0|  x_s \rangle    \\
		\langle x_s|\hat x_0 |x_0 \rangle = x_0  \langle x_0|  x_s \rangle
		\end{cases}
	\end{aligned} \quad \quad \quad
	\begin{aligned}
		\begin{cases}
			\langle y_0|\hat y_s |y_s \rangle = y_s  \langle y_0|  y_s \rangle    \\
		\langle y_s|\hat y_0 |y_0 \rangle = y_0  \langle y_0|  y_s \rangle
		\end{cases}
	\end{aligned} \eqno{(51)}
\end{equation*}
Using $(44)$ and $(46)$, the kernel also satisfies:
 \[ \text{}
	\left\{ \begin{lgathered}
	 	(x_0\cos(\epsilon s) - i \mu \frac{\partial}{\partial x_0 } \sin(\epsilon s))  \langle x_0| x_s \rangle = x_s  \langle x_0|  x_s \rangle    \\
		(x_s\cos(\epsilon s) + i \mu \frac{\partial}{\partial x_s } \sin(\epsilon s))  \langle x_s| x_0 \rangle = x_0  \langle x_s|  x_0 \rangle
	\end{lgathered} \right.\eqno{(52)}
   \] 
   
    \[ \text{}
	\left\{ \begin{lgathered}
	 	(y_0\cos(\epsilon s) + i \mu \frac{\partial}{\partial y_0 } \sin(\epsilon s))  \langle y_0| y_s \rangle = y_s  \langle y_0|  y_s \rangle    \\
		(y_s\cos(\epsilon s) - i \mu \frac{\partial}{\partial y_s } \sin(\epsilon s))  \langle y_s| y_0 \rangle = y_0  \langle x_s|  x_0 \rangle
	\end{lgathered} \right.\eqno{(53)}
   \] 
Solving $(52)$ and $(53)$, we have the kernels
 \begin{equation*}
\begin{cases}
\begin{aligned}
	\langle x_0 | x_s \rangle &= exp\left\{i\left(-\frac{1}{2 \mu \tan(\epsilon s)}(x_0^2 + x_s^2)+\frac{x_s x_0}{\mu \sin(\epsilon s)} \right)\right\}\\
	 \langle y_0 | y_s \rangle &= exp\left\{i\left(\frac{1}{2 \mu \tan(\epsilon s)}(y_0^2 + y_s^2)-\frac{y_s y_0}{\mu \sin(\epsilon s)} \right)\right\}\end{aligned}
\end{cases} \eqno{(54)}
\end{equation*} 
Using $(54)$, we switch $(\hat x_s, \hat y_s)$-representation to the $(\hat x_0, \hat y_0)$-representation:
 \begin{equation*}
   	\begin{aligned}
   		\psi_{s1}(x_0) & =\int \langle x_0| x_s \rangle \langle x_s | \psi_{s1} \rangle dx_s =  e^{- in \epsilon (-s)} \psi_0(x_0)
   	\end{aligned} \eqno{(55)}
   \end{equation*}  
    \begin{equation*}
   	\begin{aligned}
   		\psi_{s2}(y_0) & = \int \langle y_0| y_s \rangle \langle y_s | \psi_{s2} \rangle dy_s=  e^{in \epsilon (-s)} \psi_0(y_0)
   	\end{aligned} \eqno{(56)}
   \end{equation*}
where the details of the integral are in Appendix $A$.  Thus $\psi_s(x_s, y_s)$ in $(\hat x_0, \hat y_0)$-representation is:
   \begin{equation*}
	\begin{aligned}
		\psi_s(x_0,y_0) &= \psi_{s1}(x_0) \psi_{s2}(y_0) = e^{- in \epsilon (-s)} \psi_0(x_0) e^{in \epsilon (-s)} \psi_0(y_0)\\
		&=\psi_{0}(x_0, y_0)
		\end{aligned} \eqno{(57)}
\end{equation*}
There is no time dependent phase  when we switch among representations in this relativistic model. 

\subsubsection{Example 3: Bianchi Type I Universe}
Our third relativistic example is the Bianchi Type I Universe. Bianchi Type I Universe is a homogeneous cosmological solution of the Einstein field equation, which is specified by the metric \cite{2007}

\begin{equation*}
	\begin{aligned}
		ds^2 = -N^2(t)dt^2 + e^{2x(t)}e^{2\beta_{ij}(t)}dx^idx^j
	\end{aligned}
\end{equation*}
with

\begin{equation*}
	\begin{aligned}
		\beta_{ij} = \text{diag} (y + \sqrt{3} z, y -\sqrt{3} z, -2y).
	\end{aligned}
\end{equation*} Here $N(t)$ is the lapse function, $e^{2x}$ is the scale factor and $\beta_{ij}$ gives the anisotropic parameters $y(t)$ and $z(t)$. Given the Einstein-Hilbert action \cite{2007}:
\begin{equation*}
	\begin{aligned}
		\mathcal{S} = \int d^4x \sqrt{-g} (\mathcal{R} - \Lambda)
	\end{aligned}
\end{equation*} where $\mathcal{R}$ is the scalar curvature and $\Lambda$ is the cosmological constant, we have the Lagrangian of this model as \cite{2007}:

\begin{equation*}
	\begin{aligned}
		\mathcal{L} = (6 e^{3x}/N)(-\dot x^2 + \dot y^2 + \dot z^2) +Ne^{3x}\Lambda
	\end{aligned} \eqno{(58)}
\end{equation*} If we choose the gauge $N=6e^{3x}$ and set cosmological constant to 0,  the Lagrangian $(58)$ simplifies to

\begin{equation*}
	\begin{aligned}
		\mathcal{L} = -\dot x^2 + \dot y^2 + \dot z^2	\end{aligned} \eqno{(59)}
\end{equation*}
Using the formalism $(3)$ we have equations of motions and solutions

	\begin{equation*}
\begin{aligned}
	\begin{cases}
	\ddot x = 0\\

	\ddot y = 0\\
	\ddot z  = 0\\
	\end{cases}	
	\end{aligned}
	\quad \quad \quad 
	\begin{gathered}
	\begin{cases}
	 x_t= v_{x_0}t + x_0\\

	 y_t= v_{y_0}t + y_0 \\
	 
	  z_t= v_{z_0}t + z_0
	  
	\end{cases}
	\end{gathered}\eqno{(60)}
\end{equation*}
and the potential $1-$form is:

\begin{equation*}
	\theta_t(\delta q) = -v_{x_t} \delta x_t + v_{y_t} \delta y_t + v_{z_t} \delta z_t
\end{equation*}
from which we have the covariant sysmplectic $2-$form is
\begin{equation*}
	\begin{aligned}
		\omega(\delta_1 q, \delta_2 q) &=-(\delta_1v_{x_s} \delta_2 x_s - \delta_2 v_{x_s} \delta_1 x_s ) + (\delta_1v_{y_s} \delta_2 y_s - \delta_2 v_{y_s} \delta_1 y_s) + (\delta_1v_{z_s} \delta_2 z_s - \delta_2 v_{z_s} \delta_1 z_s)  \\
	\end{aligned} 
\end{equation*}
Thus the Poisson structure is

\begin{equation*}
	\begin{aligned}
		\begin{cases}
			\{x_0, v_{x_0}\} = -1 \\
			\{y_0, v_{y_0}\} =  1 \\
			\{z_0, v_{z_0}\} = 1 
		\end{cases}
	\end{aligned} \quad \quad \quad \quad
	\begin{gathered}
		\begin{cases}
			 \{x_s, v_{x_s}\} = -1 \\
			\{y_s, v_{y_s}\} = 1 \\
			\{z_s, v_{z_s}\} = 1
		\end{cases}
	\end{gathered}
\end{equation*}
Letting $(x_t, v_{x_t}, y_t, v_{y_t}, z_t, v_{z_t})$ be operators, we have the commutation relations

\begin{equation*}
	\begin{aligned}
		\begin{cases}
			[ \hat x_0, \hat v_{x_0}] = -i \\
			[ \hat y_0, \hat v_{y_0}] =  i \\
			[ \hat z_0, \hat v_{z_0}] = i 
		\end{cases}
	\end{aligned} \quad \quad \quad \quad
	\begin{gathered}
		\begin{cases}
			[ \hat x_t, \hat v_{x_t}] = -i \\
			[ \hat y_t, \hat v_{y_t}] =  i \\
			[ \hat z_t, \hat v_{z_t}] = i 
		\end{cases}
	\end{gathered} \eqno{(61)}
\end{equation*}
The Hamiltonian constraint of this model is

\begin{equation*}
	\begin{aligned}
		\mathcal{H} &= \dot x \frac{\partial \mathcal{L} }{\partial \dot x} + \dot y \frac{\partial \mathcal{L} }{\partial \dot y} + \dot z \frac{\partial \mathcal{L} }{\partial \dot z} - \mathcal{L} = -v_{x_t}^2 + v_{y_t}^2 + v_{z_t}^2 
	\end{aligned} \eqno{(62)}
\end{equation*}
In $(\hat x_s, \hat y_s, \hat z_s)$-representation, we have

\begin{equation*}
	\begin{aligned}
		\langle x_s, y_s, z_s |\mathcal{H} | \psi \rangle &= \langle x_s, y_s, z_s | (-\hat v_{x_s}^2 + \hat v_{y_s}^2 + \hat v_{z_s}^2)| \psi \rangle \\
		 &= (\frac{\partial^2}{\partial x_s^2 } - \frac{\partial^2}{\partial y_s^2 } -\frac{\partial^2}{\partial z_s^2 } )\psi_s(x_s, y_s, z_s) =0
	\end{aligned} \eqno{(63)}
\end{equation*}
Letting $\psi_s(x_s, y_s, z_s) = \psi_{s1}(x_s)\psi_{s2}(y_s)\psi_{s3}(z_s) $, we have
\begin{equation*}
	\begin{aligned}
		\begin{cases}
			\frac{\partial^2}{\partial x_s^2 } \psi_{s1}(x_s) = - \eta_1^2 \psi_1(x_s)\\
			\frac{\partial^2}{\partial y_s^2 } \psi_{s2}(y_s) = - \eta_2^2 \psi_1(y_s)\\
			\frac{\partial^2}{\partial z_s^2 } \psi_{s3}(z_s) = - \eta_3^2 \psi_1(z_s)
		\end{cases}
	\end{aligned}
	\begin{gathered}
		\begin{cases}
			\psi_{s1}(x_s) = \mathcal{C}_1 e^{i\eta_1 x_s} + \mathcal{D}_1 e^{-i\eta_1 x_s}\\
			\psi_{s2}(y_s) = \mathcal{C}_2 e^{i\eta_2 y_s} + \mathcal{D}_2 e^{-i\eta_2 y_s}\\
			\psi_{s3}(z_s) = \mathcal{C}_3 e^{i\eta_3 z_s} + \mathcal{D}_3 e^{-i\eta_3 z_s}
		\end{cases}
	\end{gathered} \eqno{(64)}
\end{equation*} 
where $\eta_i\ (i = 1, 2, 3)$ are the separation constants that satisfy the condition that $\eta_1 ^2 = \eta_2^2 + \eta_3^2 $. Similarily we have the eigenstates  of the Hamiltonian in the representation $\hat x_0, \hat y_0, \hat z_0$ as $\psi_0(x_0, y_0, z_0) = \psi_{1}(x_0)\psi_{2}(y_0)\psi_{3}(z_0)$ with
\begin{equation*}
	\begin{aligned}
		\begin{cases}
			\frac{\partial^2}{\partial x_0^2 } \psi_{1}(x_0) = - \eta_1^2 \psi_1(x_0)\\
			\frac{\partial^2}{\partial y_0^2 } \psi_{2}(y_0) = - \eta_2^2 \psi_1(y_0)\\
			\frac{\partial^2}{\partial z_0^2 } \psi_{3}(z_0) = - \eta_3^2 \psi_1(z_0)
		\end{cases}
	\end{aligned}
	\begin{gathered}
		\begin{cases}
			\psi_{1}(x_0) = \mathcal{C}_1 e^{i\eta_1 x_0} + \mathcal{D}_1 e^{-i\eta_1 x_0}\\
			\psi_{2}(y_0) = \mathcal{C}_2 e^{i\eta_2 y_0} + \mathcal{D}_2 e^{-i\eta_2 y_0}\\
			\psi_{3}(z_0) = \mathcal{C}_3 e^{i\eta_3 z_0} + \mathcal{D}_3 e^{-i\eta_3 z_0}
		\end{cases}
	\end{gathered} \eqno{(65)}
\end{equation*} 
with $\eta_1 ^2 = \eta_2^2 + \eta_3^2 $. Using $(60)$ and $(61)$ we want the kernel to satisfy

\begin{equation*}
	\begin{aligned}
		\begin{cases}
			\langle x_0 | \hat v_{x_0}s + \hat x_s | x_s \rangle = x_s \langle x_0 | x_s \rangle \\
			\langle x_s | -\hat v_{x_s}s + \hat x_0 | x_0 \rangle = x_0 \langle x_0 | x_s \rangle
		\end{cases}
	\end{aligned}
	\begin{gathered}
		\begin{cases}
			(-i\frac{\partial }{\partial x_0}s + x_0) \langle x_0 | x_s \rangle = x_s \langle x_0 | x_s \rangle  \\
			(i\frac{\partial }{\partial x_s}s + x_s) \langle x_s | x_0 \rangle = x_0 \langle x_s | x_0 \rangle  \\

		\end{cases}
	\end{gathered} 
\end{equation*}

\begin{equation*}
	\begin{aligned}
		\begin{cases}
			\langle y_0 | \hat v_{y_0}s + \hat y_s | y_s \rangle = y_s \langle y_0 | y_s \rangle \\
			\langle y_s | -\hat v_{y_s}s + \hat y_0 | y_0 \rangle = y_0 \langle y_0 | y_s \rangle
		\end{cases}
	\end{aligned}
	\begin{gathered}
		\begin{cases}
			(i\frac{\partial }{\partial y_0}s + y_0) \langle y_0 | y_s \rangle = y_s \langle y_0 | y_s \rangle  \\
			(-i\frac{\partial }{\partial y_s}s + y_s) \langle y_s | y_0 \rangle = y_0 \langle y_s | y_0 \rangle  \\

		\end{cases}
	\end{gathered}
\end{equation*}

\begin{equation*}
	\begin{aligned}
		\begin{cases}
			\langle z_0 | \hat v_{x_0}s + \hat x_s | x_s \rangle = x_s \langle x_0 | x_s \rangle \\
			\langle x_s | -\hat v_{x_s}s + \hat x_0 | x_0 \rangle = x_0 \langle x_0 | x_s \rangle
		\end{cases}
	\end{aligned}
	\begin{gathered}
		\begin{cases}
			(i\frac{\partial }{\partial z_0}s + z_0) \langle z_0 | z_s \rangle = z_s \langle z_0 | z_s \rangle  \\
			(-i\frac{\partial }{\partial z_s}s + z_s) \langle z_s | z_0 \rangle = z_0 \langle z_s | z_0 \rangle  \\

		\end{cases}
	\end{gathered}
\end{equation*} 
Solving these, we have 

\begin{equation*}
	\begin{aligned}
		\begin{cases}
			\langle x_0 | x_s \rangle = e^{i(-\frac{1}{2s}(x_0^2 + x_s^2) + \frac{x_0x_s}{s} )}\\
			\langle y_0 | y_s \rangle = e^{i(\frac{1}{2s}(y_0^2 + y_s^2) - \frac{y_0y_s}{s} )}\\
			\langle z_0 | z_s \rangle = e^{i(\frac{1}{2s}(z_0^2 + z_s^2) -\frac{z_0z_s}{s} )}
		\end{cases}
	\end{aligned} \eqno{(66)}
\end{equation*} Using $(50)$ we switch from $(\hat x_0, \hat y_0)$-representation to $(\hat x_s, \hat y_s)$-representation:
\begin{equation*}
	\begin{aligned}
		\psi_{01}(x_s) &= \int \langle x_s |x_0 \rangle \langle x_0 | \psi_{01} \rangle = e^{-i\frac{s}{2} \eta_1^2}\psi_{s1}(x_s)		
	\end{aligned}
\end{equation*}
\begin{equation*}
	\begin{aligned}
		\psi_{02}(y_s) &= \int \langle y_s |y_0 \rangle \langle y_0 | \psi_{02} \rangle= e^{i\frac{s}{2} \eta_2^2}\psi_{s2}(y_s)		
	\end{aligned}
\end{equation*}

\begin{equation*}
	\begin{aligned}
		\psi_{03}(z_s) &= \int \langle z_s |z_0 \rangle \langle z_0 | \psi_{03} \rangle= e^{i\frac{s}{2} \eta_3^2}\psi_{s3}(z_s)		
	\end{aligned}
\end{equation*}
Combining these we have $\psi_0(x_0, y_0, z_0)$ in the $(\hat x_s, \hat y_s, \hat z_s)$-represenation:
\begin{equation*}
\begin{aligned}
\psi_0(x_s, y_s, z_s) &= \psi_{01}(x_s) \psi_{02}(y_s) \psi_{03}(z_s)=e^{-i\frac{s}{2} \eta_1^2}\psi_{01}(x_s) e^{i\frac{s}{2} \eta_2^2}\psi_{02}(y_s)e^{i\frac{s}{2} \eta_3^2}\psi_{03}(z_s)\\
                     &= e^{-i\frac{s}{2}( \eta_1^2 -\eta_2^2 -\eta_3^2) } \psi_{01}(x_s)\psi_{02}(y_s)\psi_{03}(z_s)\\
                     &=\psi_s(x_s)\psi_s(y_s)\psi_s(z_s)
\end{aligned} \eqno{(67)}
\end{equation*}
with $ \eta_1^2 = \eta_2^2 +\eta_3^2$. As in the previous examples, the time dependent phase cancels when we switch among representations, indicating that the quantum states again do not evolve in coordinate time $t$.

\section{Summary and Discussion} 

In this article we explored  some quantum aspect of covariant phase space. The covariant phase space is the manifold of the space of solutions, equipped with a symplectic covariant $2-$form. There is no time parameter $t$ in the covariant phase space, as each point of the space of solutions is a whole solution of the equations of motion---a point is a whole dynamical trajectory of the system. When the space of solutions is uniquely determined by the values of the solutions at a specific moment, these values of solutions can be considered as coordinates. Classically, the time evolution of the system in covariant phase space is described by the the change of the coordinates of the manifold of the space of solutions. The equations of motion determine the transition functions of these coordinates. 

To quantize covariant phase space, we treat these coordinates of the manifold of the space of solutions as operators, the coordinate operators. The eigenstates of these operators are used as a basis of the Hilbert space, giving a specific representation. Here we proposed that the time evolution of the quantum states of the covariant phase space is restored by performing the transformation of the states among these representations. This proposition is shown by comparing the results of quantization for a non-relativistic model (the harmonic oscillator) and three relativistic models (cosmological models). We notice that: $1)$ In our non-relativistic model, when the states are transformed among different representations, a time dependent phase appears, which is exactly the produced out by the time evolution operator acting on the stationary states in the usual quantization method. $2)$ In relativistic models, no such time dependent phase emerges, indicating that the quantum states in these models do not evolve along a particular coordinate time $t$. This is consistent with other quantization method applied to in these models \cite{PhysRevD.77.044023, Bina:2007wj, 2007}. 
 
 This difference between the non-relativistic model and relativistic 
models here may come from the different roles of their Hamiltonians and 
time parameters. In a non-relativistic model, the Hamiltonian operator 
generates time evolution, and it has non-vanishing eigenvalues. In 
relativistic models, on the other hand, the Hamiltonian acts as a 
constraint, with eigenvalue zero. Time in a non-relativistic model is 
unique and absolute. It is natural to see the spatial variables in these 
non-relativistic models, and thus the quantum states, evolve with 
respect to this unique time.  In cosmological models, on the other hand, 
the time variable is merely a time coordinate. This time coordinate is 
entangled with the spatial coordinates, and its choice is rather 
arbitrary and nonunique. The lack of evolution of the quantum states in 
such a time coordinate reflects the fact that in general relativity a 
particular time coordinate is only an arbitrary internal variable, on 
the same footing with spatial coordinates, with no external physical 
significance. It would be interesting to see whether this proposition 
holds when this method applied to more general models.

Instead of using the time coordinate that corresponds to a Hamiltonion 
constraint, there have been attempts to use the the matter fields 
coupled with gravity as references for dynamical evolution.  For 
instance, in \cite{doi:10.1142/S0218271811019347} the scalar field in a 
cosmological model is considered as an ``internal clock'', with respect 
to which the quantum states evolve. It would be interesting to see how 
this scenario can be applied in the quantization method proposed in this 
article.

\section{Appendix}
   \subsection{Appendix A: Integrals}
   In this section, we exhibit the details of the integrals that involve the Hermite functions in $(55)$, $(56)$ and $(17)$:
   \begin{equation*}
	\begin{aligned}
		\begin{cases}
			\int dx_s \left \langle x_0 | x_s \right \rangle e^{-\frac{x_s^{2}}{2 \eta}}H_{n}(\frac{x_s}{\sqrt{\eta}} ) &= \int dx_s e ^{(-i)(\alpha x_s^{2}+\beta x_sx_0 + \gamma x_1^{2} )} e^{-\frac{x_s^{2}}{2 \eta}}H_{n}(\frac{x_s}{\sqrt{\eta}} )\\
			\int dy_s \left \langle y_0 | y_s \right \rangle e^{-\frac{y_s^{2}}{2 \eta}}H_{n}(\frac{y_s}{\sqrt{\eta}} ) &= \int dy_s e ^{(+i)(\alpha y_s^{2}+\beta y_sy_0 + \gamma y_0^{2} )} e^{-\frac{y_s^{2}}{2 \eta}}H_{n}(\frac{y_s}{\sqrt{\eta}} )
		\end{cases}
	\end{aligned}
\end{equation*}with
\begin{equation*}
	\begin{aligned}
		\alpha = \frac{\cos(\epsilon s )}{2 \eta \sin(\epsilon s)},\quad \quad \beta = -\frac{1}{ \eta \sin (\epsilon s)}, \quad \quad \gamma = \frac{\cos(\epsilon s )}{2 \eta \sin(\epsilon s)}
	\end{aligned}
\end{equation*}The technique is similar to that used in doing the Fourier transformation of the Hermite functions. Here we have applied two formulas. The first one is the generating function of the Hermite functions \cite{andrews1998special}:
\begin{equation*}
	\begin{aligned}
		e^{-\frac{1}{2}z^2+2z\rho-\rho^2} = \sum_{n=0}^{\infty} e^{-\frac{1}{2}z^{2}} H_{n}(z)\frac{\rho^{n}}{n!}
	\end{aligned} \eqno{(68)}
\end{equation*} 
The second is \cite{doi:10.1080/00029890.1998.12004954}:

   \begin{equation*}
	\begin{aligned}
		\int _{-\infty}^{+\infty}dx e^{ax(x-2b)}=\ (\pm i) \frac{\sqrt{\pi}}{\sqrt{a}}e^{-ab^2}
	\end{aligned} \eqno{(69)}
\end{equation*}where $a, b \in\mathbb{C}$ with $re(a)<0$. Letting $x_s=\sqrt{\eta} z$, we do the integral of the product of the left side of $(68)$ with  $e ^{(-i)(\alpha x_s^{2}+\beta x_sx_0 + \gamma x_0^{2} )}$ which gives:
\begin{equation*}
	\begin{aligned}
		&\int dx_s e^{-\frac{1}{2}z^2+2z\rho-\rho^2}e ^{(-i)(\alpha x_s^{2}+\beta x_sx_0 + \gamma x_0^{2} )}
		=\sqrt{\eta}\int dz e^{-\frac{1}{2}z^2+2z\rho-\rho^2}e ^{(-i)(\alpha \eta z^{2}+\beta \sqrt{\eta}z x_0 + \gamma x_0^{2} )}\\
		&=\sqrt{\eta}e^{-i \gamma x_0^2} e^{- \rho^2}\int dz e^{(-i \alpha \eta- \frac{1}{2})z^2+(2 \rho -i\beta \sqrt{\eta}x_0)z}\\
		&=\mathcal{C}e^{-i \gamma x_0^2}  \exp{(-e^{2i\epsilon s} \rho^2 +\frac{2}{\sqrt{\eta}}e^{i \epsilon s}x_0 \rho +\frac{ie^{i \epsilon s}}{2 \eta sin(\epsilon s)}x_0^2)}
	\end{aligned} \eqno{(70)}
\end{equation*}
  Next let $\tau = e^{i\epsilon s} \rho, x_0=\sqrt{\eta} \xi_1$. The $(63)$ becomes:
  \begin{equation*}
	\begin{aligned}
		&\int dx_s  \exp{(-\frac{1}{2}z^2+2z\rho-\rho^2)} \exp{((-i)(\alpha x_s^{2}+\beta x_sx_0 + \alpha x_0^{2} ))}\\ 
		&= \mathcal{C} \exp{(-i\gamma x_0^2 )} \exp{(-\tau^2+2\tau \xi_1+i\frac{e^{i\epsilon s }}{2sin(\epsilon s)} \xi_1^2)}\\
		&= \mathcal{C} e^{-i\gamma x_0^2}e^{\frac{i}{2} \frac{ \cos(\epsilon s)}{\sin(\epsilon s)}\xi_1^2} \sum_{n=0}^{\infty}e^{-\frac{1}{2}\xi_1^2}H_{n}(\xi_1)\frac{\tau^n}{n!}\\
		&= \mathcal{C}  \sum_{n=0}^{\infty}e^{-\frac{1}{2 \eta}x_0^2}H_{n}(\frac{1}{ \sqrt{\eta}} x_0)\frac{e^{in\epsilon s } \rho^n}{n!}
		\end{aligned} \eqno{(71)}
\end{equation*} Doing the integral of the right side of $(61)$ with $e^{((-i)(\alpha x_s^{2}+\beta x_sx_0 + \gamma x_0^{2} ))}$,
\begin{equation*}
	\begin{aligned}
		&\sum_{n=0}^{\infty} \int dx_s \exp{((-i)(\alpha x_s^{2}+\beta x_sx_0 + \alpha x_0^{2}))} \exp{(-\frac{1}{2}z^2)} H_n(z) \frac{\rho^{n}}{n!} \\
		&=\sum_{n=0}^{\infty} \int dx_s \exp{((-i)(\alpha x_s^{2}+\beta x_sx_0 + \alpha x_0^{2}))} e^{-\frac{1}{2\eta}x_0^2} H_n(\frac{1}{\sqrt{\eta}}x_0) \frac{\rho^{n}}{n!}
	\end{aligned} \eqno{(72)}
\end{equation*} Identifying the term with the same order of $\rho$ in $(71)$, $(72)$, we have
\begin{equation*}
	\begin{aligned}
		 \int dx_s \left \langle x_0 | x_s \right \rangle e^{-\frac{x_s^{2}}{2 \eta}}H_{n}(\frac{x_s}{\sqrt{\eta}} ) &=\int dx_s e ^{(-i)(\alpha x_s^{2}+\beta x_sx_0 + \gamma x_0^{2} )} e^{-\frac{x_s^{2}}{2 \eta}}H_{n}(\frac{x_s}{\sqrt{\eta}} ) \\ &=\mathcal{C}  e^{in \epsilon t} e^{-\frac{1}{2} (\frac{1}{\sqrt{\eta}} x_0)^2} H_n(\frac{1}{ \sqrt{\eta}}x_0^{2})
	\end{aligned} 
\end{equation*}
Similarly,

\begin{equation*}
	\begin{aligned}
		 \int dy_s \left \langle y_0 | y_s \right \rangle e^{-\frac{y_s^{2}}{2 \eta}}H_{n}(\frac{y_s}{\sqrt{\eta}} ) &=\int dy_s e ^{i(\alpha y_s^{2}+\beta y_sy_0 + \gamma y_0^{2} )} e^{-\frac{y_s^{2}}{2 \eta}}H_{n}(\frac{y_s}{\sqrt{\eta}} ) \\ &=\mathcal{C}  e^{-in \epsilon t} e^{-\frac{1}{2} (\frac{1}{\sqrt{\eta}} y_0)^2} H_n(\frac{1}{ \sqrt{\eta}}y_0^{2})
	\end{aligned} \eqno{(73)}
\end{equation*} as used in previous sections.

\subsection{Appendix B}
 Here we exhibit the details of the integrals involving with the Airy functions in $(39)$, $(40)$ and $(41)$. To get $(39)$, $(40)$, we have
\begin{equation*}
\begin{aligned}
	\psi_{s1} (v_{x_0})&=\int \langle v_{x_0} | x_s \rangle \langle x_s | \psi_{s1} \rangle dx_s 
	=\int  e^{i(\frac{1}{2}v_{x_0}^2s + v_{x_0}(\frac{1}{2}V_0s^2 - x_s) + \frac{1}{8}V_0^2s^3 - sV_0x_s )} \psi_{s1}(x_s)d x_s\\
		&= \int e^{i(\frac{1}{2}v_{x_0}^2s + v_{x_0}(\frac{1}{2}V_0s^2 - x_s) + \frac{1}{8}V_0^2s^3 - sV_0x_s )}Ai(\frac{\eta -2V_0 x_s}{(2V_0)^{2/3}})dx_s\\
		&= e^{i(\frac{1}{2} v_{x_0}^2s+v_{x_0}(\frac{1}{2}V_0s^2) + \frac{1}{8} V_0^2s^3 +(-v_{x_0}-V_0s) \eta (2V_0)^{-1} )}\int e^{i(v_{x_0}+V_0s)(2V_0)^{-\frac{1}{3}}x_{\eta}} Ai(x_{\eta})dx_{\eta} \\
		&= \mathcal{C}e^{i(-\frac{1}{6V_0}v_{x_0}^3 -\frac{\eta v_{x_0}}{2V_0})}e^{i(-\frac{1}{24}V_0^2s^3- \frac{s}{2} \eta )}  
\end{aligned} \eqno{(39)}
\end{equation*}

\begin{equation*}
\begin{aligned}
\psi_{s2}(v_{y_0}) &= \int \langle v_{y_0} | y_s \rangle \langle y_s | \psi_{s2} \rangle dy_s =\int e^{-i(\frac{1}{2}v_{y_0}^2s + v_{y_0}(\frac{1}{2}V_0s^2 - y_s) + \frac{1}{8}V_0^2s^3 -s V_0y_s )} \psi_{s2} dy_s \\
		&= \int e^{-i(\frac{1}{2}v_{y_0}^2s + v_{y_0}(\frac{1}{2}V_0s^2 - y_s) + \frac{1}{8}V_0^2t^3 - sV_0y_s )}Ai(\frac{\eta -2V_0 y_s}{(2V_0)^{2/3}})dy_s\\
		&= e^{-i(\frac{1}{2} v_{y_0}^2s+v_{y_0}(\frac{1}{2}V_0s^2) + \frac{1}{8} V_0^2s^3 +(-v_{y_0}-V_0s) \eta (2V_0)^{-1} )}\int e^{-i(v_{y_0}+V_0s)(2V_0)^{-\frac{1}{3}}x_{\eta}} Ai(y_{\eta})dy_{\eta} \\
				&=\mathcal{C} e^{i(\frac{1}{6V_0}v_{y_0}^3 +\frac{\eta v_{x_0}}{2V_0})}e^{i(\frac{1}{24}V_0^2s^3+ \frac{s}{2} \eta )}
	\end{aligned} \eqno{(40)}
\end{equation*}
where $x_{\eta} = \frac{\eta -2V_0 x_s}{(2V_0)^{2/3}}$ $y_{\eta} = \frac{\eta -2V_0 y_s}{(2V_0)^{2/3}}$ and $\mathcal{C}$ is a constant fixed by the normalization condition. Here we have used the formula for the Airy function:

\begin{equation*}
 \mathcal T(Ai)(k) := \int_{- \infty }^{+\infty } Ai(x) e^{-2 \pi i k x}dx= e^{\frac{i}{3}(2\pi k)^3 }	
\end{equation*}
To get $(41)$ we have:
\begin{equation*}
	\begin{aligned}
		\psi_s(v_{x_0},v_{y_0}) 
		&= \psi_{s1}(v_{x_0}) \psi_{s2}(v_{y_0}) \\
		&=  e^{i(-\frac{1}{6V_0}v_{x_0}^3 -\frac{\eta v_{x_0}}{2V_0})}e^{i(-\frac{1}{24}V_0^2s^3- \frac{s}{2} \eta )} \times e^{i(\frac{1}{6V_0}v_{y_0}^3 +\frac{\eta v_{y_0}}{2V_0})}e^{i(\frac{1}{24}V_0^2s^3+ \frac{s}{2} \eta )}\\
		&= e^{i(-\frac{1}{6V_0} v_{x_0}^3 -\frac{1}{2V_0} \eta v_{x_0})}  e^{i(\frac{1}{6V_0} v_{y_0}^3 + \frac{1}{2V_0} \eta v_{y_0})} \\
		&= \psi_{v_0} (v_{x_0}, v_{y_0})
	\end{aligned} \eqno{(41)}
\end{equation*}

\section*{Acknowledgments}
I want to thank Steve Carlip for valuable discussions and comments.
   This work was supported in part by Department of Energy grant
   DE-FG02-91ER40674.
\bibliographystyle{unsrt}  
\bibliography{ref}

\end{document}